# Sick of being driven? – Prevalence and modulating factors of carsickness in the European population in context of automated driving

Myriam Metzulat[a,b], Barbara Metz[c], Aaron Edelmann[d], Alexandra Neukum[e], Wilfried Kunde[f]

[a] Julius Maximilians University Würzburg, Department of Psychology III, Röntgenring 11, 97070 Würzburg, Corresponding author: myriam.metzulat@stud-mail.uni-wuerzburg.de **and**
[b] WIVW GmbH; Robert-Bosch-Straße 4, 97209 Veitshöchheim, Germany, Corresponding author: metzulat@wivw.de
[c] WIVW GmbH; Robert-Bosch-Straße 4, 97209 Veitshöchheim, Germany, metz@wivw.de
[d] AUDI AG; Auto-Union-Straße 1, 85045 Ingolstadt, Germany, aaron.edelmann@audi.de
[e] WIVW GmbH; Robert-Bosch-Straße 4, 97209 Veitshöchheim, Germany, neukum@wivw.de
[f] Julius Maximilians University Würzburg, Department of Psychology III, Röntgenring 11, 97070 Würzburg, wilfried.kunde@uni-wuerzburg.de

*Abstract* – As in automated driving the driver becomes a passenger, carsickness might reduce comfort for susceptible individuals. Insights in the prevalence of carsickness and its modulating factors are considered useful for the development of automated vehicles to mitigate or prevent its occurrence. An online survey was conducted with $N$ = 3999 participants in Spain, Sweden, Poland, and Germany. 30% of participants reported to have already experienced carsickness as adult. The frequency of carsickness was modulated not only by demographic factors (country, gender, age), but also by frequency of being a passenger, type of non-driving related task, road type, and the seating position in car. Furthermore, the efficiency of applied countermeasures, temporal aspects of carsickness development, as well as the relation of carsickness with the acceptability of automated driving and the effect on subjective fitness to drive was investigated. The results are discussed with focus on automated driving.

## 1. Introduction

One of the major benefits of automated driving (AD) is that the occupant can freely use the time saved during automation due to the relief of the driver's role. However, engaging in non-driving related tasks (NDRTs) causes symptoms of motion sickness in susceptible individuals (Diels & Bos, 2016). Motion sickness is likely to affect the occupant's comfort. Comfort can be described as "the subjective feeling of pleasantness of driving/riding in a vehicle in the absence of both physiological and psychological stress" (Carsten & Martens, 2019). It is also considered to be an aspect or outcome of successfully applied ergonomics. Ergonomics considers the reduction or avoidance of any kind of physical or mental strain when interacting with the environment (Bridger, 2008). As motion sickness can make AD a rather uncomfortable experience, thus reducing its benefits, it is of interest to assess whether motion sickness is a frequent problem for car occupants and to know its modulating factors. This will indicate whether motion sickness will be a challenge for the successful introduction of AD and what aspects need to be considered in its design.

Motion sickness includes symptoms like drowsiness, dizziness, headache, sweating, salivation to nausea and, in extreme cases, vomiting (Graybiel et al., 1968; Reason & Brand, 1975). It results from conflicting sensory inputs, i.e. the vestibular and visual systems, according to the theory of sensory conflict (Reason & Brand, 1975). Motion sickness occurs in a variety of different modes of transport, such as ships, trains or road vehicles. The latter is known as carsickness. It is a known phenomenon among passengers, while drivers are usually not affected (Dong et al., 2011; Rolnick & Lubow, 1991). Here, the mismatch between expected and perceived motion due to the lack of control over the vehicle as a passenger contributes to the development of carsickness (Reason, 1978).

### 1.1. Prevalence and modulating factors of motion and carsickness

There have been a few studies on the prevalence of motion sickness or carsickness, reporting an average prevalence of around 50%. In a large scale survey conducted by Turner and Griffin (1999), 58% of respondents reported having experienced some form of motion sickness with carsickness being the most frequent one (37%). However, the sample might not be representative for the general population as it included only relatively young coach travelers. The same applies to the sample of Reason and Brand (1975), which included only students. 2/3 of them reported carsickness as passengers at some point in their lives. Schmidt et al. (2020) only included participants who regularly use public transport and/ or private cars, as these groups are particularly at risk of carsickness. 46% of participants reported experiencing some degree of carsickness as a car passenger in the past five years.

Studies investigating the prevalence of carsickness frequently ask if respondents ever experienced symptoms either during a specified time span or life time. The frequency and severity of symptoms and accompanying factors are mostly not considered in detail. As potentially modulating factors, age and gender are often investigated. Various studies have found, that women are significantly more susceptible to motion sickness than men (e.g., Klosterhalfen et al., 2005; Schmidt et al., 2020) and that the susceptibility in adults decreases with age (e.g., Bos et al., 2007; Schmidt et al., 2020). However, other factors may also be relevant to carsickness, such as the frequency of driving or being a passenger. Situational factors that influence the frequency and severity of carsickness are also of interest. These may be characteristics of the driving situation, (e.g. road type, driving style etc.), but also specific passenger behaviors that may promote or hinder the occurrence of carsickness. As far as the driving situation is concerned, many turns as well as curvy roads are strongly associated with carsickness occurrence (Schmidt et al., 2020). Experimental studies show that the engagement in NDRTs, especially visual ones, induce carsickness to a significant extent (Brietzke et al., 2021; Jones et al., 2019; Metzulat et al., 2024). Subjective reports on the relation between NDRTs and carsickness reveal that reading and using a device are the most provocative ones (Diederichs et al., 2024; Schmidt et al., 2020).

### 1.2. Carsickness countermeasures and effects for automated driving

In terms of countermeasures against carsickness, the most common responses in an international survey were stopping to read or using a device (Diederichs et al., 2024). There are also many experimental studies, which test specific countermeasures. Visual or haptic cues and motion planning are the most studied ones in context of automated driving (Pereira et al., 2024). Further concepts include pleasant olfactory cues, ventilation, pleasant music, breathing exercises or a combination of several methods (D'Amour et al., 2017; Emond et al., 2025; Peck et al., 2020; Stromberg et al., 2015). While experimental studies systematically test specific countermeasures, self-reported effectiveness of different countermeasures is usually based on more than one use and could be therefore more reliable. Furthermore, mitigating carsickness might even be safety relevant as there might be a negative impact of carsickness on driving performance (Metzulat et al., 2025). As long as the automated vehicle (AV) is not fully autonomous, the driver needs to be able to take vehicle control back safely when the AV reaches its system boundaries. If carsickness reduces the ability to safely steer a car and to react appropriately to affordances of the situation, carsickness might lead to reduced takeover performance. As carsickness might decrease comfort and be a safety concern, it is likely, that it will compromise the acceptability of AD in susceptible individuals. Therefore, it is of special interest if there is a relation between the proneness for carsickness and the acceptability of AD.

## 2. Objectives

Our aim was to address the topic of carsickness in the context of AD from a broader perspective. The survey assesses the prevalence of carsickness but also the impact of modulating factors that might be relevant for the development of AD functions. The prevalence of carsickness amongst the European population was assessed including participants from central, northern, southern, and eastern Europe. The focus was on being a passenger on the front seat and engaging in other activities, as this is the most relevant situation for AD. Going beyond previous studies, a wide range of potentially modulating factors was included, like personal factors (e.g., driving experience) as well as situational factors (driving situations or NDRTs). Temporal aspects and the efficiency of applied countermeasures are also of interest. Insights in factors influencing the occurrence and severity of carsickness might be useful to design automated functions such that the occurrence of carsickness is minimized or prevented. Furthermore, topics related to AD, which might be affected by carsickness were addressed, i.e. the acceptability of AD functions and the impact on the subjective fitness to drive.

## 3. Materials and methods

To reach as many participants we conducted an online survey. As the study was part of the European project Hi-Drive (www.hi-drive.eu) the aim was to include respondents from different parts of Europe. We therefore included participants from Germany, Spain, Sweden, and Poland. The questionnaire was first implemented in German and English and then translated to Spanish, Swedish and Polish.

### 3.1. Questionnaire

The survey started with a welcome message explaining the purpose of the survey as well as an informed consent. At the beginning of the questionnaire all respondents filled in items assessing demographics, driving experience and information on the frequency of use of different modes of transport. To ensure a common understanding of motion sickness among participants a definition of motion sickness was provided for the respective questions. The English definition and the translation used for each language can be found in the appendix.

To assess the relation of carsickness with the acceptability of AD functions, questions on the willingness to use AD functions and the likelihood of engaging in NDRTs while these functions are active followed. The following AD functions were included: driving on a motorway, in a traffic jam, and in urban traffic.

Next, the Motion Sickness Susceptibility Questionnaire Short (MSSQ; Golding, 1998) was included to assess overall motion sickness susceptibility as well as a simple yes/no question if one has ever experienced motion sickness in any type of transportation as an adult.

For measuring susceptibility to carsickness, a yes/no question was asked as to whether the respondent had ever experienced carsickness as a passenger in a car in adult life. Only if participants answered with yes, more detailed questions about carsickness experiences followed including items on potential modulating factors like the frequency of carsickness as a passenger, when doing different NDRTs, e.g. reading, and in different driving situations, e.g. curves, motorway, etc. These questions were assessed on an 8-point categorical scale with 5-points covering the frequency between never and always, plus extra categories for the answers 'not yet experienced', 'don't know', and 'I actively avoid this situation to prevent sickness'. Furthermore, the severity of carsickness symptoms had to be assessed on the Misery Scale (MISC; Bos et al., 2005) for the different modes of transport.

To capture the temporal aspects of carsickness, participants were asked about the amount of time after which symptoms typically occur when engaging in other activities as a car passenger. We also asked how long the symptoms lasted after the car journey.

To find out which are the most common and successfully used countermeasures and prevention strategies for carsickness, we asked participants to indicate on a list of countermeasures whether they had ever used them and, if so, whether they had helped.

To assess if carsickness might be a safety issue, we asked participants to imagine driving with symptoms of carsickness and to answer whether they would feel fit to drive, feel impaired by carsickness and avoid certain actions due to carsickness, e.g. turning their head. If there is interest in the full questionnaire, the authors can make it available on request.

### 3.2. Participants and data collection

Participants were recruited by Kantar, a market research institute offering online surveys, and to a small extent ($n$ = 47) via the test-driver panel of the Würzburg Institute for Traffic Sciences (WIVW). Invitations were sent to people who were at least 18 years old (age of adulthood and age at which you can drive a car on your own with a driving license). The survey included quotas filled to have equal numbers per gender and age group in each country. Data collection took place from the 13th to 26th of April 2022 and from 4th to the 21st of July 2022. Participants recruited by Kantar received financial compensation for completing the survey. $N$ = 4667 participants completed the survey. After a quality check, $N$ = 668 had to be excluded due to inconsistencies. Thus, $N$ = 3999 valid completions remained. The average completion time was $M$ = 11 minutes. Table 1 shows the sample description across all participants and separated by country.

**Table 1:** *Sample description per country*

|  | Germany | Poland | Sweden | Spain | Total |
|---|---|---|---|---|---|
| N | 1102 | 959 | 971 | 967 | 3999 |
| Age | $M$ = 47.6, | $M$ = 45.2, | $M$ = 49.9, | $M$ = 46.2, | $M$ = 47.2, |
|  | $SD$ = 18.4 | $SD$ = 17.1 | $SD$ = 18.0 | $SD$ = 17.2 | $SD$ = 17.8 |
| Gender | 548 female | 478 female | 553 female | 495 female | 2074 female |
|  | 552 male | 481 male | 418 male | 469 male | 1920 male |
|  | 2 diverse | 0 diverse | 0 diverse | 3 diverse | 5 diverse |
| % drivers' licence | 85% | 81% | 81% | 87% | 84% |
| Years of driving experience | $M$ = 26.1 | $M$ = 26.1 | $M$ = 30.5 | $M$ = 23.4 | $M$ = 24.8 |
|  | $SD$ = 17.5 | $SD$ = 17.5 | $SD$ = 18.1 | $SD$ = 16.1 | $SD$ = 17.2 |
| Mileage last 12 months (km) | $M$ = 10 159 | $M$ = 13 826 | $M$ = 6 970 | $M$ = 10 272 | $M$ = 9 971 |
|  | $SD$ = 16 716 | $SD$ = 27 530 | $SD$ = 18 145 | $SD$ = 17 541 | $SD$ = 18 166 |
| % car owner | 74% | 76% | 67% | 83% | 75% |

### 3.3. Statistical Analysis

For the statistical analysis of the frequency of carsickness items, only the 5-point scale responses from never to always on the 8-point scale were included, while the categories 'never experienced', 'don't know' and 'I avoid this situation to prevent carsickness' were excluded. This was done to obtain an ordinal scale. Due to the scale level, non-parametric test methods were chosen for the items using this scale. For those participants who answered 'no' to the question whether they had ever experienced carsickness as an adult ($N$ = 2815), the response category 0 = never was used for the frequency items. The use of specific analysis methods is described in the relevant results section. A significance level of $\alpha$ = .05 was applied.

## 4. Results

### 4.1. Overall prevalence and demographics

Overall, about 29.6% of all participants reported that they had at least experienced carsickness once in their adult life. This number increases to 40.7% when motion sickness in any kind of transport or environments is concerned. Participants reporting that they had already experienced carsickness also reported a significantly higher MSSQ raw value compared to those who had not [$t(3972)=$-50.119, $p <$.001]. Table 2 shows the mean MSSQ raw values and percentages of carsickness and motion sickness experience as an adult person.

**Table 2:** *MSSQ Raw values as well as percentage of motion sickness and carsickness experience in adult life*

| Measure | Germany | Poland | Sweden | Spain | Total |
|---|---|---|---|---|---|
| MSSQ Raw | $M$ = 10.67, $SD$ = 12.66 $N$ = 1083 | $M$ = 11.54, $SD$ = 13.74 $N$ = 957 | $M$ = 12.88, $SD$ = 12.87 $N$ = 970 | $M$ = 14.50, $SD$ = 12.80 $N$ = 964 | $M$ = 12.35, $SD$ = 13.13 $N$ = 3974 |
| % motion sickness | 31.9% | 35.8% | 44.1% | 52.2% | 40.7% |
| % carsickness | 23.0% | 23.0% | 33.6% | 39.7% | 29.6% |

To see the effects of gender, age, and country on the incidence of carsickness a binary logistic regression model was calculated including these factors as predictors. The model predicted the likelihood of having already experienced carsickness significantly [$\chi^2(8)$ = 644.514, p < .001; total adjusted $R^2$ = 13.26%]. Country [$\chi^2(3)$ = 98.41, p < .001], gender [$\chi^2(1)$ = 258.74, p < .001] and age group [$\chi^2(3)$ = 278.65, p < .001] were all significant predictors. The odds ratios for each level of the predictors are shown in Table 3. For each predictor the group with the lowest likelihood was chosen as the reference. Spanish people have a more than two times higher chance of experiencing carsickness than people in Poland and women have a more than three times higher chance than men. Only German participants do not differ significantly to Polish people having a similar carsickness incidence. All age groups differ significantly to the reference group of 65+ with the youngest age group having a more than six times higher chance. Figure 1 displays the carsickness incidence by country, gender, and age.

**Table 3:** *Odds ratios for the binary logistic regression for the likelihood of experiencing carsickness (N = 3994)*

| Predictor | Level | Odds Ratio | Lower 95% CI | Upper 95% CI |
|---|---|---|---|---|
| Country | *Poland (Ref)* | *1.00* | - | - |
| | Germany | 1.05 | 0.84 | 1.30 |
| | Sweden | 1.86 | 1.50 | 2.32 |
| | Spain | 2.48 | 2.00 | 3.06 |
| Gender | *Male (Ref)* | *1.00* | - | |
| | Female | 3.10 | 2.66 | 3.62 |
| Age | *65+ (Ref)* | *1.00* | - | - |
| | 46 - 65 | 2.87 | 2.24 | 3.71 |
| | 31 - 45 | 4.87 | 3.80 | 6.28 |
| | 18 - 30 | 6.37 | 4.96 | 8.23 |

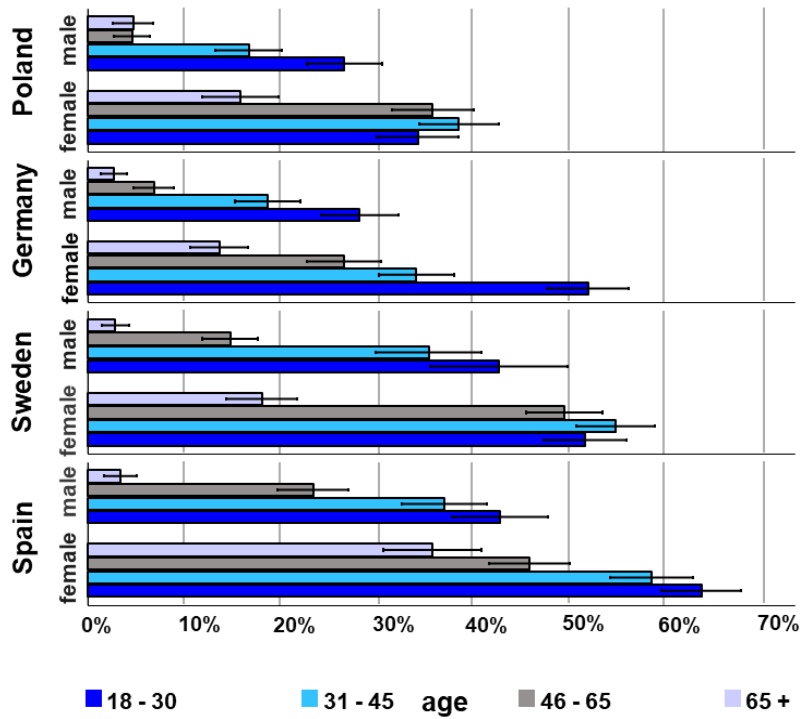

**Fig. 1:** Prevalence of Carsickness by country, gender (without the $n$ = 5 diverse; $N$ = 3994), and age. Error bars represent the 95% CIs.

### 4.2. Modulating factors
#### 4.2.1. Frequency of transport modes and driving experience

Figure 2 shows the frequency of use for the different transport modes which was collected as potentially modulating factor. Being a car driver and a car passenger on the front seat where the most frequent travel modes. 47.7% of participants drive a car (almost) every day and 39.9% are front passengers every day or 1 − 2 times a week.

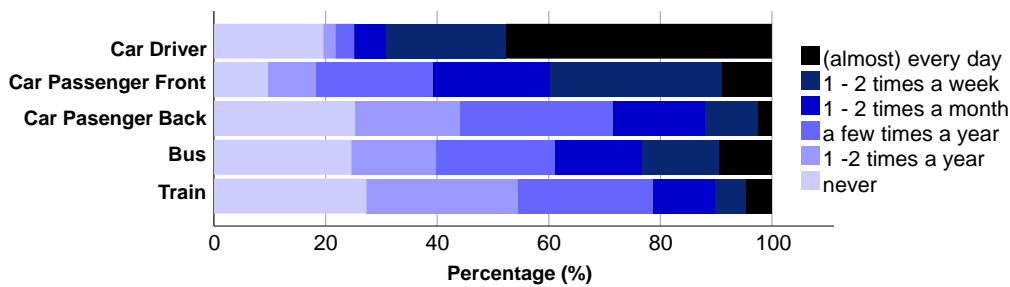

**Fig. 2:** Frequency of travelling in different transport modes ($N$ = 3999).

$N$ = 1184 respondents stated that they had already experienced carsickness as an adult and were therefore asked all the detailed carsickness questions. Most of these participants (75%) stated, that they rarely or sometimes experience carsickness. Only a small proportion (2.5%) reported to always experience carsickness as a front passenger. Figure 3 shows the frequencies of carsickness experience for different positions in car. Most of the car drivers never get sick (56%). 75% stated, that they rarely or sometimes experience carsickness as a front passenger. On the back seat the biggest proportion was the category "sometimes to often" (66%) as in backwards direction (46%).

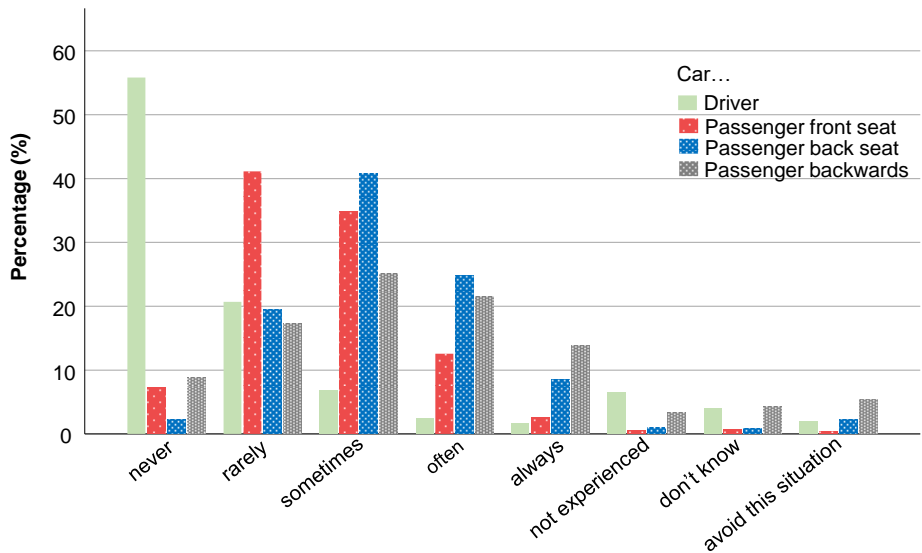

**Fig. 3** Frequency of carsickness experience per position in car (*N* = 1184)

Corresponding to the reported frequencies, the severity of symptoms assessed based on MISC differs in the repeated measure one-way ANOVA significantly between modes of transport, [$F_{(4.195, 7248.767)}$ = 471.290, p < .001]. Post hoc tests show that all modes of transport differ significantly from each other, except for being a passenger in a car travelling backwards and being a passenger in a bus. The strongest symptoms occur when being a backseat passenger (see Figure 4).

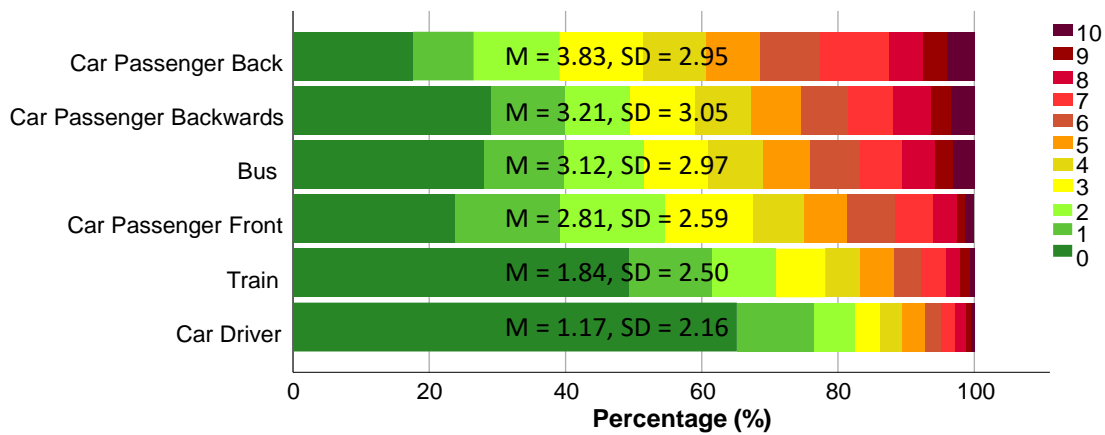

**Fig. 4** Carsickness severity (MISC; Bos et al., 2005) for different modes of transport with descriptive values (*N* = 1184)

Looking in more detail into the reported frequency of different transport modes, it is noticeable that women are much more frequent front passengers than men (see grey areas in Table 4). It is therefore conceivable that the gender effect reported in section 4.1 is due to the frequency of being a passenger and thus the possibility of experiencing carsickness in the first place. To test this assumption, the data were analyzed using a hierarchical procedure. As outcome variable, the frequency of carsickness as a front passenger (from 0 = never to 4 = always) was used. First, an ordinal logistic regression model was calculated with age and frequency of being a front passenger as predictors. In the second model, gender was additionally included as a predictor. The two models were statistically compared to see if gender still contributed significantly to the prediction when frequency of being a passenger was already included.

**Table 4:** *Number of participants per frequency of being a passenger on the co-driver seat.*

|  |  | never | 1-2 times a year | few times a year | 1-2 times a Month | 1-2 times a week | (almost) every day |
|---|---|---|---|---|---|---|---|
| Gender | female | 122 | 128 | 362 | 436 | 798 | 228 |
|  | Male | 258 | 223 | 474 | 398 | 436 | 131 |
| total |  | 380 | 351 | 836 | 834 | 1234 | 359 |

In both models, the frequency of being a front passenger is a significant predictor with all predictor levels differing significantly to the reference of never being a passenger (see 95% CIs of odds ratios in Table 5). The second model with the additional predictor of gender significantly improved the model prediction [LR(1) = 180.40, $p$ < .001], leading to the conclusion that gender is a significant predictor in addition to the effect of frequency of being a passenger (women 2.83 times more likely than men to experience carsickness more frequently).

**Table 5:** *Ordinal regression models for frequency of carsickness as passenger with and without the predictor Gender (N = 3994).*

|  |  | Without Gender | | | With Gender | | |
|---|---|---|---|---|---|---|---|
| Predictor | Level | Odds Ratio | Lower 95% CI | Upper 95% CI | Odds Ratio | Lower 95% CI | Upper 95% CI |
| Age | *65+ (Ref)* | *1.00* | - | - | *1.00* | - | - |
|  | 46 - 65 | 2.88 | 2.24 | 3.71 | 2.91 | 2.25 | 3.80 |
|  | 31 - 45 | 4.33 | 3.80 | 6.28 | 4.33 | 3.35 | 5.63 |
|  | 18 - 30 | 4.84 | 4.96 | 8.23 | 4.95 | 3.83 | 6.46 |
| Frequency of being a passenger | *Never (Ref)* | *1.00* | - | - | *1.00* | - | - |
|  | 1-2 times (year) | 1.82 | 1.16 | 2.88 | 1.77 | 1.12 | 2.81 |
|  | Few times (year) | 2.35 | 1.62 | 3.50 | 2.12 | 1.45 | 3.18 |
|  | 1-2 times (Month) | 3.06 | 2.11 | 4.53 | 2.54 | 1.74 | 3.79 |
|  | 1-2 times week | 3.62 | 2.54 | 5.32 | 2.72 | 1.89 | 4.02 |
|  | almost every day | 3.10 | 2.07 | 4.75 | 2.36 | 1.56 | 3.64 |
| Gender | *Male (Ref)* | - | - | - | *1.00* | - | - |
|  | Female | - | - | - | 2.83 | 2.42 | 3.31 |

As age and driving experience are correlated strongly ($r$ = .869, $p$ < .001, $N$ = 3330) partial Spearman correlations were computed between the driving experience in years and the frequency of experiencing carsickness as a front passenger, while controlling for the confounding effect of age. This resulted in a small but significant negative correlation of $r$ = -.058 ($p$ = .002, $N$ = 3330).

### 4.2.2. Non-driving related tasks and driving situations

To test the effect of different NDRTs and driving conditions on the frequency of carsickness, a Kruskal-Wallis H-test for independent measures was calculated as a dependent comparison method would have required too many subjects to be excluded, as for many participants at least for one item the answer was one of the excluded answer categories, e.g. 'don't know'.

There was a strong impact of NDRT on the carsickness frequency ($H$(9, $N$= 9980) =1412.5, $p$ < .001). Figure 5 shows the carsickness frequencies for each NDRT ordered from highest prevalence to lowest prevalence. *Reading* was rated as the task, which induced carsickness the most frequent (48.1% often – always) followed by *smartphone use* (45.4% often – always). The tasks with the highest proportion of participants saying they actively avoid them to prevent carsickness are using a *notebook* (14.4%), *video* *watching on an integrated display* (12.9%) and *video on a notebook* (12.8%), followed by *reading* (7.8%) and *using a smartphone* (5%).

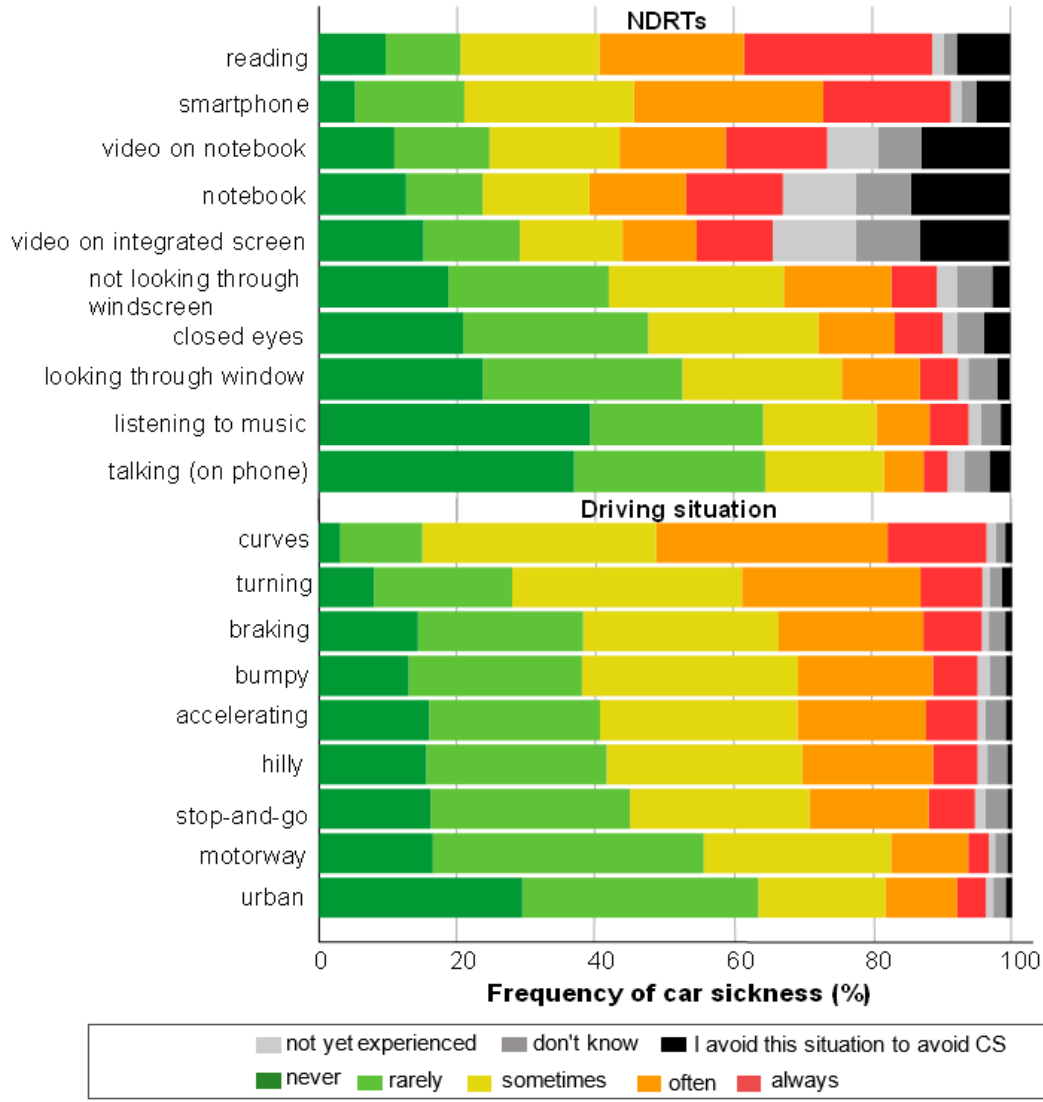

**Fig. 5** Frequencies of carsickness (%) for different NDRTs and driving situations (*N* = 1184)

Furthermore, there was also a significant impact of the driving situation on the frequency of carsickness (*H*(8, *N* = 10183) =872.7, *p* < .001). *Curves* were reported as causing carsickness more frequently than other driving situations, followed by *turning*. *Urban driving* and *motorway* were associated with the lowest frequencies of carsickness, while all other situations ranged between those two extremes, see Fig. 5

### 4.3. Temporal aspects of carsickness

Table 6 shows the onset time of carsickness symptoms when being a car passenger and engaging in other activities as well as the duration of symptoms after stopping the drive. About a third experience symptoms immediately or within less than 5 minutes and about the same proportion experience symptoms lasting at least 10 – 30 minutes.

**Table 6:** *Proportion (%) of answers (N = 1184) for the onset time and duration time of symptoms after the drive.*

|  | Immediately there | < 5min | 5 – 10 min | 10 -30 min | > 30 min | No symptoms | Don't know | Avoid this situation |
|---|---|---|---|---|---|---|---|---|
| **Onset time** | 11.3 % | 22.6% | 22.0% | 22.0% | 9.9% | 19% | 4.1% | 6.4% |
|  | Immediately gone | - | 5 – 10 min | 10 - 30 min | 30 – 60 min | > 60 min | Don't know | - |
| **Duration** | 9.6% | - | 24.7% | 30.0% | 13.6% | 9.0% | 13.1% | - |

### 4.4. Countermeasures against carsickness

Regarding potential countermeasures against carsickness, participants could indicate whether they had already used a particular countermeasure and that it had worked or that it had not worked, or that they had never tried it. In Figure 6, the responses to the countermeasure questions are ordered by the proportion of participants who said that a countermeasure had worked. The most effective countermeasure was *stopping secondary tasks (*72%). The countermeasures *choose seat in front row* (68%), *have a break* (67%), *open window* (67%) and *become a driver* (54%) were all successfully applied by more than half of the participants. Further commonly used countermeasures that were also often applied successfully were related to redirecting gaze (through window/on horizon) or lowering the temperature/air conditioning.

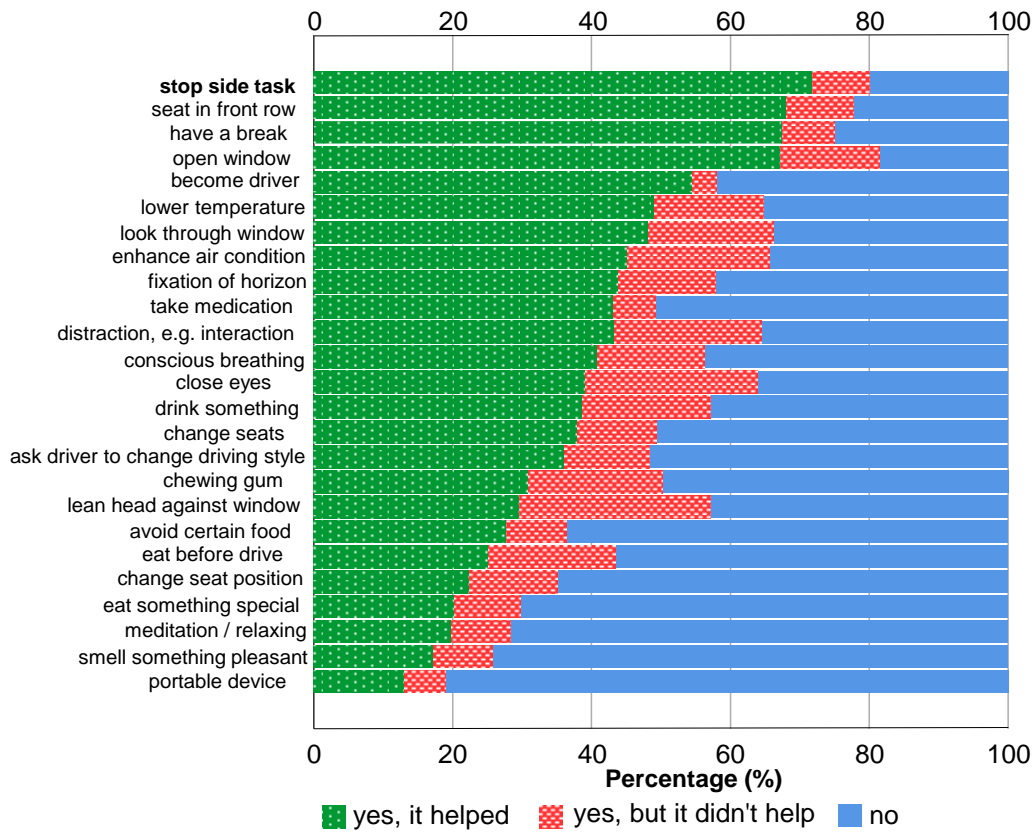

**Fig. 6** Answers regarding countermeasures use: percentage of already applied and helped, applied and didn't help or not applied (*N* = 1184)

### 4.5. Driving under the influence of carsickness

The proportion of answers regarding the experience of driving under the influence of carsickness can be found in Table 7. More than a third of participants would not feel fit to drive and about 45% would feel impaired with carsickness.

**Table 7:** *Percentage of answers to questions regarding driving under the influence of carsickness (N = 3999)*

| Answer | With symptoms of carsickness would you… | | |
| | … feel fit to drive? | … think your driving behavior is impaired or changed? | … avoid certain actions while driving?* |
| --- | --- | --- | --- |
| yes | 24.6 % | 44.5 % | 20.8% |
| no | 35.8% | 23.0% | 37.0% |
| don't know | 39.6% | 32.5% | 42.2% |

*such as turn your head or look down on the speedometer

### 4.6. Acceptability of AD and carsickness

There were only small but significant ($p$ <.001) positive Spearman correlations between the frequency of experiencing carsickness and the willingness to use AD functions ($r$ = .098, $N$ = 3979) and to engage in NDRTs when these functions are active ($r$ = .102, $N$ = 3738). For the analysis, ratings of the 3 automated functions (motorway, traffic jam and urban chauffeur) were averaged. The Kruskal-Wallis-Test showed a significant impact of the frequency of experiencing carsickness as a front passenger on the willingness to use these functions ($H$(4, $N$ = 3979) = 39.6, $p$ < .001) and to engage in NDRTs ($H$(4, $N$ = 3738) = 40.6, $p$ < .001). Figure 7 shows the averaged values for willingness to use and to engage in NDRTs for the different carsickness frequencies.

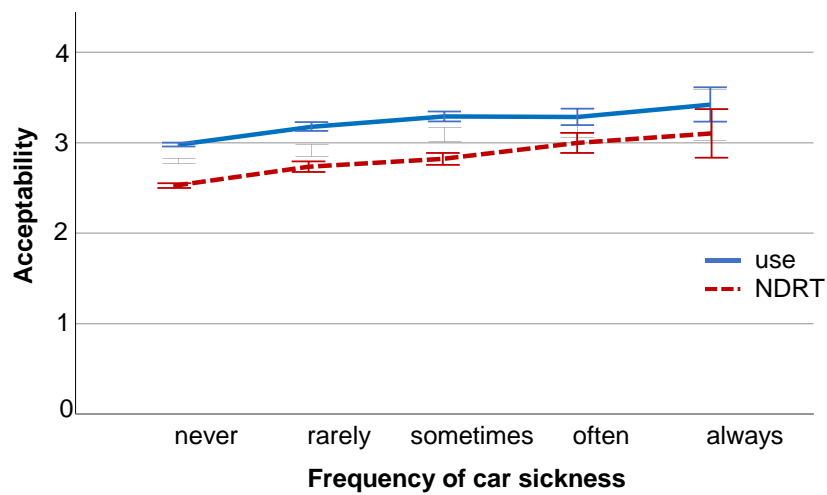

**Fig. 7:** Acceptability of AD functions and engagement in NDRTs for carsickness frequencies as a front passenger: Willingness to use from 0 = very unwillingly to 5 = very much and likelihood to engage in NDRTs during automation from 0 = not at all to 5 = very likely.

## 5. Summary and discussion

### 5.1. Prevalence and modulating factors of carsickness

In our sample about 30% of respondents reported having experienced carsickness in their adult life, when motion sickness in any kind of transportation is included, it increases to 41%. This is a lower prevalence than reported in other studies (e.g. Schmidt et al. 2020, Turner & Griffin 1999, Reason & Brand, 1975.) However, the variation in the exact wording of the question used for assessing the prevalence (e.g. relation to different time spans) and the variation in selection of the sample (e.g. in Schmidt et al. 2020 regularly public transport and car users) makes a direct comparison of values difficult. As our results show that the frequency of being a passenger significantly predicts the carsickness incidence, this probably explains why the presented prevalence is lower compared to Schmidt et al. (2020). The results of Schmidt et al. (2020) may therefore overestimate the prevalence in the general population only focusing on those who frequently travel. However, those excluded, could still use AD functions in the future.

Furthermore, we found a significant difference in the prevalence between countries. Spain shows almost double the prevalence of Germany and Poland with Sweden in between. One possible explanation could be either differences in language, meaning that the term for carsickness is associated differently, e.g. with a different symptom severity. For this reason, however, we had included a definition with example symptoms so that the same understanding of the term should be given. Another explanation could be cultural differences in reporting the experience of negative corporal symptoms. A study on cross-country differences in self-reported health supports this assumption, revealing country differences in reporting styles, with Spanish people reporting the poorest health (Jürges, 2007). It is also conceivable that carsickness differences are due to differences between countries regarding road types, road conditions or driving styles. If we assume – what is reasonable and shown by our data – that there is an impact of road conditions and the available road infrastructure on the frequency of carsickness, it can also be assumed that there should be differences between regions. In this case, a comparison between studies with the aim of deriving some kind of baseline prevalence for carsickness might not be meaningful as regional characteristics can be expected to impact the results. Future research could look at the effect of, for example, proportion of curvy and hilly roads versus straight and flat roads in a person's usual driving area on carsickness prevalences or even make comparisons between countries with the proportions of such roads.

Our results replicate the frequently reported effect of age and gender as modulating factors (see [1.1](#)), but it also shows that additional factors more related to living circumstances are relevant as well. The frequency of being a passenger is a significant factor predicting the reported likelihood of experiencing carsickness and reducing the gender effect slightly but not significantly. Furthermore, there are differences between transportation modes and seating positions. Travelling on the back seat or in backwards direction in car lead most frequently and most severely to carsickness. Driving experience showed a small but negative relation to carsickness, meaning that the more driving experience you have the less carsick you get. Although the effect was very small, it seems plausible that the more accustomed you are to driving, the better you are at judging vehicle movements probably also as a passenger, and the less likely you are to get sick. In addition, the driving situations play a crucial role in sickness development with driving maneuvers with higher lateral dynamics leading to more carsickness, as found in Schmidt et al. (2020). These results highlight that the reported prevalence is not simply based on specific characteristics of a person but results from an interplay between relevant personal characteristics (e.g. age), aspects of current living conditions (e.g. having a driver's license) and characteristics of the living environment (e.g. road network, topography of the residential area).

### 5.2. Carsickness and automated driving

The results show that in AD the prevalence of carsickness can be expected to be highest in situations with high lateral accelerations but rather low during ordinary highway driving with comparably little changes of acceleration. As the latter is likely to be the most common situations in the early stages of adoption, it is unlikely that there are many cases of carsickness in the initial period. As far as NDRT engagement is concerned, the visual tasks reading as well as using devices like smartphones or laptops, were reported to be the most provocative activities, which is consistent with the finding from other surveys (Diederichs et al., 2024; Schmidt et al., 2020). These tasks are therefore considered most relevant activities for occupant monitoring systems to detect (Diederichs et al., 2024) and hence for anti-motion sickness systems to address in AVs. Stopping the visual NDRTs was the most effective countermeasure. This is not a surprise; however, it limits the potential benefit of automation for users having problems with carsickness. Further countermeasure that helped many (>50%), were sitting in front row, having a break, opening a window, or becoming a driver. Most of them are rather irrelevant for anti-carsickness programs in AVs and some would even interrupt the automation. Interestingly, smelling something pleasant, which so far has been investigated in several studies as potential mitigation program in AD (Emond et al., 2025), was never used by most of our sample and for those who did use it, it was not successful for some.

The possibility of spending the time in the car instead on driving on other freely chosen activities is promoted as one of the main benefits of AD. Unexpectedly, there is no strong relation between the proneness for carsickness and the reported acceptability of AD. Drivers reporting experiencing carsickness more often, show even higher acceptability for AD. One possible explanation would be that this is modulated by driving frequency. Driving more frequently might be linked to experiencing carsickness more often but also to perceiving automation as more useful as it might be beneficial during daily life. Furthermore, susceptible people may have higher expectations that AD could solve their problem through a more comfortable and anticipatory driving styles. More susceptible people were also more willing to engage in NDRTs during automation, which also seems contradictory. However, those, who are generally more willing to engage in NDRTs consequently experience carsickness more often, but are also willing to engage in NDRTs during automation as they normally do as passengers.

More than half of people experience symptoms within 5-10 minutes, which is quite fast considering a longer AD journey. In addition, around 45% of people experienced symptoms for at least 10 minutes, and some for more than an hour. We also found that 36% would not feel fit to drive and 45% would feel impaired while driving due to carsickness, which is in line with objective results on cognitive impairment relevant to driving (Metzulat et al., 2025). The long-lasting symptoms combined with the associated subjectively reduced driving ability represent a potential safety risk for AD when taking over and subsequently driving under the influence of carsickness. This could be the case either when a system limit is reached, e.g. in an emergency where a rapid and appropriate response is required, or when the driver wants to take over to reduce symptoms. As becoming a driver has been reported as one of the most helpful countermeasures, taking over is likely to be a promising way to alleviate carsickness symptoms for a significant proportion of users, making research on takeovers and driving with carsickness even more relevant.

### 5.3. Methodological considerations and limitation

While questionnaires are a practical, efficient, and economical tool to reach many participants, self-reported data might be biased by specific response styles reducing the validity of responses, e.g. socially desirable or extreme responding (Paulhus & Vazire, 2007). Response styles exist not only on a personal but also on a cultural level (Jürges, 2007), as discussed above concerning the cross-country differences. Furthermore, self-reported data might be affected by poor memory, misinterpretation of questions or especially in context of motion sickness the familiarity with concepts e.g., travelling with different transport modes (Hartmann et al., 2025). Nevertheless, they are widely accepted as the method of choice for motion sickness prevalence studies, especially for cross-national studies.

As mentioned, the inclusion criteria as well as the wording of the question including the related time span for the carsickness experience influence the prevalence substantially and complicate comparisons between studies. This should be kept in mind, when interpreting differences in found prevalences.

When recruiting susceptible participants for experimental studies, e.g. to test countermeasures where certain levels of carsickness need to be induced, we recommend asking about susceptibility in situations that resemble the experimental situation closely, e.g. car passenger watching a video. This should improve the estimation of actual carsickness levels in the study, as our survey showed significant differences in prevalences between modes of transport, positions in the car as a passenger and different NDRTs.

## 6. Conclusions

Our data revealed an overall prevalence of carsickness for the adult life of 30%. Modulating factors for carsickness frequency as a front passenger, as most likely situation for AD, were gender, age, country of residence, frequency of being a passenger, driving experience and situational factors like the road type, type of NDRT and seat position. This shows, that there is an interplay between personal characteristics, living situation and environment as well as behavioral aspects. These findings should be considered when developing AVs. For example, when designing seating concepts, the finding that rear-facing seats most often cause carsickness should be taken into account. Other than expected, the carsickness frequency is not related to reduced acceptability of AD. Susceptible individuals are even more likely to use automation for NDRTs, which highlights the importance for driver monitoring systems detecting carsickness associated activities or carsickness itself combined with anti-carsickness programs to mitigate symptoms.

**Author contributions: Myriam Metzulat:** Conceptualization, Methodology, Data Curation, Formal Analysis, Investigation, Writing - Original Draft, Visualisation. **Barbara Metz:** Conceptualization, Writing - Review & Editing, Project administration. **Aaron Edelmann:** Conceptualization, Writing - Review & Editing**. Alexandra Neukum:** Funding acquisition. **Wilfried Kunde:** Writing - Review & Editing, Supervision.

**Conflicts of Interest:** The authors declare no conflict of interest.

**Acknowledgements:** This project has received funding from the European Union's Horizon 2020 research and innovation programmed under grant agreement No 101006664. The sole responsibility of this publication lies with the authors. Neither the European Commission nor CINEA – in its capacity of Granting Authority – can be made responsible for any use that may be made of the information this document contains. The authors would like to thank Dominik Mühlbacher, Christina Kremer and Markus Tomzig for their valuable feedback on the questionnaire.

**Funding:** This project was funded by the European Union's Horizon 2020 research and innovation program under grant agreement no. 101006664.

## 7. Appendix

*Definitions of motion sickness in English and used languages*

**English:** Motion sickness is characterized by a state of discomfort that can occur when travelling by ship (seasickness) but also, for example, by car, bus, or train. Symptoms vary from person to person, but often include fatigue, dizziness, headache, sweating, awareness of the stomach or nausea (in extreme cases, even vomiting). It can also include relatively mild symptoms such as slight discomfort, awareness of the stomach, faintness, head pressure, strained eyes, etc.

**German:** Reisekrankheit zeichnet sich durch einen Zustand des Unwohlseins aus, der zum Beispiel bei Fahrten mit Schiffen (Seekrankheit) aber auch Autos oder Achterbahnen auftreten kann. Die Symptome unterscheiden sich von Person zu Person, umfassen jedoch häufig Müdigkeit, Schwindel, Kopfschmerzen, Schweißausbrüche oder Übelkeit (in extremen Fällen bis hin zu Erbrechen). Hierzu zählen auch verhältnismäßig leichte Symptome wie leichtes Unwohlsein, Bemerkbar-Machen des Magens, Mattigkeit, Kopfdruck, angestrengte Augen etc.

**Polish:** Choroba lokomocyjna charakteryzuje się złym samopoczuciem, które może wystąpić np. podczas podróży statkiem (choroba morska), ale także samochodem lub kolejką górską. Objawy różnią się w zależności od osoby, ale często obejmują zmęczenie, zawroty głowy, ból głowy, pocenie się lub nudności (w tym wymioty w skrajnych przypadkach). Obejmuje to również stosunkowo łagodne objaws, takie jak nieznacznie złe samopoczucie, odczuwalny dyskomfort żołądka, omdlenia, ucisk w głowie, nadwyrężone oczy itp.

**Swedish:** Åksjuka karaktäriseras av ett tillstånd av obehag som kan inträffa när man åker båt (sjösjuka), men också, när man åker bil, buss eller tåg. Symptomen på åksjuka varierar från person till person, men innefattar ofta trötthet, yrsel, huvudvärk, svettningar, orolig mage eller illamående (i extremfall även kräkningar). Notera att definitionen även innefattar relativt milda symptom såsom lätt obehag, orolig mage, yrsel, huvudvärk, ansträngda ögon.

**Spanish:** El mareo se caracteriza por un estado de malestar que puede producirse, por ejemplo, al viajar en barco, coches o incluso en montañas rusas. Los síntomas varían de una persona a otra, pero suelen incluir fatiga, vértigo, dolores de cabeza, sudoración o náuseas (en casos extremos hasta vómitos). También se incluyen síntomas relativamente leves, como un ligero malestar, hacer notar al estómago, presión en la cabeza, tensión en los ojos, etc.